%Paper: astro-ph/9412030
%From: scheel@astrosun.TN.CORNELL.EDU (Mark Scheel)
%Date: Thu, 8 Dec 94 17:00:10 EST

\magnification=\magstep1
\baselineskip=18pt
\vbadness=2700

\def\ss {\smallskip}
\def\ms {\medskip}
\def\bs {\bigskip}
\def\ctl {\centerline}
\def\nid {\noindent}

\def\bB {\bf{B}}
\def\bF {\bf{F}}
\def\bE {\bf{E}}

\def\bJ {\bf{J}}

\def\bv {\bf{v}}

\def\cE {\cal{E}}

\def\cF {\cal{F}}

\def\bnb {\pmb{$\nabla$}}

\def\ap {\alpha}
\def\dt {\delta}
\def\Dt {\Delta}

\def\kp {\kappa}

\def\nb {\nabla}
\def\om {\omega}

\def\ov {\over}
\def\pt {\partial}

\def\sm {\sigma}

\def\los {\buildrel < \over \sim}
\def\gos {\buildrel > \over \sim}

\def\pmb#1{\setbox0=\hbox{#1}%
   \kern-.025em\copy0\kern-\wd0
   \kern.05em\copy0\kern-\wd0
   \kern-.025em\raise.0433em\box0 }

\centerline{\bf{SPIN-UP/SPIN-DOWN OF MAGNETIZED STARS}}
\centerline{\bf{WITH ACCRETION DISKS AND OUTFLOWS}}
\ms
\centerline{\it{R.V.E. Lovelace$^{1,*}$, M.M. Romanova$^2$, G.S.
 Bisnovatyi-Kogan$^2$}}
\ss
\ctl{1.~~Institute of Astronomy, Madingley Road, Cambridge CB3 0HA}

\ctl{2.~~Space Research Institute, Russian Academy of Sciences, Moscow,
Russia~~~~~~}

\bs
\ctl{\bf{ABSTRACT}}
\ms

An investigation is made of disk accretion of matter onto a rotating
star with an aligned dipole magnetic field.  A new aspect of this work
is that when the angular velocity of the star and disk differ
substantially we argue that the $\bf B$ field linking the star and disk
rapidly inflates to give regions of open field lines extending from
the polar caps of the star and from the disk.  The open field line
region of the disk leads to the possibility of magnetically driven
outflows.  An analysis is made of the outflows and their back affect on
the disk structure assuming an ``$\ap$" turbulent viscosity model for
the disk and a magnetic diffusivity comparable to this viscosity.  The
outflows  are found to extend over a range of radial distances
inward to a distance close to $r_{to}$, which is the distance of
the maximum of the angular rotation rate of the disk.  We find
that $r_{to}$  depends on the star's
magnetic moment, the accretion rate, and the disk's magnetic
diffusivity. The outflow regime is accompanied in general by a spin-up of
the rotation rate of the star.  When $r_{to}$ exceeds the star's
corotation radius $r_{cr} = (GM/\om_*^2)^{1\ov 3}$, we argue that
outflow solutions do not occur, but instead that ``magnetic braking"
of the star by the disk due to field-line twisting
occurs in the vicinity of $r_{cr}$.  The magnetic braking solutions
can give spin-up or
spin-down (or no spin change) of the star depending mainly on the star's
magnetic moment and the mass accretion rate.  For a system with $r_{to}$
comparable to $r_{cr}$, bimodal behavior is possible where extraneous
perturbations (for example, intermittency of $\ap$, $\bf B$ field flux
introduced from the companion star, or variations in the mass accretion
rate) cause the system to flip between spin-up (with outflows, $r_{to} <
r_{cr}$) and spin-down (or spin-up) (with no outflows, $r_{to} > r_{cr}$).

\medskip
\nid $^*$ On leave from Department of Applied Physics, Cornell University,
Ithaca, NY 14853.

\vfill\eject
\nid {\bf{I.~~INTRODUCTION}}
\ss

The problem of matter accretion on to magnetized stars has been of
continued interest since the discovery of X-ray pulsars in binary
systems (Giacconi et al. 1971; Schreier et al. 1972; Tananbaum et al.
1972). The X-ray pulsars were interpreted as magnetized neutron stars,
accreting matter from a companion star (Pringle and Rees 1972; Davidson
and Ostriker 1973; Lamb, Pethick, and Pines 1973; Rappaport and Joss
1977). Of the more than 40 of known X-ray pulsars, most are in high-mass
($M_{tot}>15 $) binary systems, while only few are in low-mass
($M_{tot} < 3 $) binary systems (see, for example, review by Nagase
1989).

From the early observations, many X-ray pulsars were found to show
secular decreases of pulse period (spin-up of the neutron star
rotation) with relatively large rates of change of $-\dot
P/P=10^{-2}-10^{-6}~{\rm yr}^{-1}$ (Gursky and Schreier 1975; Schreier
and Fabbiano 1976; Rappaport and Joss 1983). The long-term monitoring of
X-ray pulsars during the last two decades have shown a wide variety of
pulse-period evolution on different time-scales. Some X-ray pulsars show
a steady spin-up evolution during many years which rapidly changes to a
steady spin-down evolution (e.g., GX 1+4, SMC X-1, 4U 1626-67). Others
show the wavy variations of $P$ on time-scales of a few years on a
background of systematic spin-up or spin-down (e.g., Cen X-3, Her X-1,
Vela X-1) (Nagase 1989; Sheffer et al. 1992; Bildsten 1993).

Short-term fluctuations of pulse-period on time-scales of days to
months, including clear evidence of spin-down episodes, were found for a
number of sources, for example, Her X-1 (Giacconi 1974), Cen X-3
(Fabbiano and Schreier 1977), Vela X-1 (Nagase et al. 1984). Detailed
observations of Vela X-1 with Hakucho and Tenma satellites, and with
HEAO-1 have revealed fluctuations of large amplitude with
time-scales about
3-10 days, which were estimated to be random in both amplitude and sign
(Nagase 1981). BATSE observations have shown that in some cases (Cen
X-3, Her X-1) the timescale of fluctuations is less than one day with
the change of sign from spin-up to spin-down and vice-versa (Bildsten
1993). Note that some X-ray pulsars (usually the long-period ones) show
no evidence of regular spin-up or spin-down (for example, 4U 0115+63, GX
301-2).

In some cases, like Her X-1, the existence of an accretion disk is
indicated from various observations in the X-ray and optical bands (for
example, Middleditch and Nelson 1976; Bisnovatyi-Kogan et al. 1977;
Middleditch, Puetter, and Pennypacker 1985). In several other cases,
such as Cen X-3, SMC
X-1, GX 1+4, and 4U 1626-67, there are also different kinds of evidence,
that matter from companion star forms a disk (for example, Tjemkes et
al. 1986; Nagase 1989). Most of these are short-period pulsars with
period about several seconds or less (excluding GX 1+4 with period 122
s.). However, most of long-period pulsars, are supposed to be powered by
the capture of stellar wind from companion. However if the companion is
the star of Be type with slow rate of outflow, then it is probable that
once again the disk first forms, opposite to the case when the companion
star is of O-B kind with high speed of matter outflow (see review of
Nagase 1989). Illustrative data for two pulsars are shown in Figure 1.

Some aspects of matter accretion by magnetized neutron stars were
considered for the first time by Bisnovatyi-Kogan and Friedman (1970),
and by Shvartsman (1971). Ideas on the spinning-up of pulsars by
accretion were first proposed in the case of disk-fed pulsars by Pringle
and Rees (1972), Lamb, Pethick and Pines (1973), and Lynden-Bell and
Pringle (1974), and for wind-fed
pulsars by Davidson and Ostriker (1973), and Illarionov and Sunyaev (1975)
(see also Lipunov 1993).  The spin-down of a rapidly rotating
magnetized star due to ``propellar" action of the magnetosphere
was suggested by Shvartsman (1971) and Illarinov and Sunyaev (1975),
but a detailed model has not been worked out.
Pringle and Rees (1972) considered matter
accretion from a viscous disk to a magnetized rotating neutron star. Their
theory considered the idealized case of an aligned rotator where the
magnetic moment of pulsar is aligned parallel or anti-parallel with the
pulsar's spin-axis and the normal to the plane of the accretion disk.
They supposed that accreting matter stops at the point where the
magnetic pressure of the magnetosphere becomes equal to pressure of
matter. At this distance, matter transfers its angular momentum to the
star through a thin boundary layer. Another idea regarding the width of
the transition zone between the unperturbed accretion disk and
magnetosphere, was considered in series of papers by Ghosh and Lamb
(1978, 1979a,b) (hereafter denoted GL). They assumed that magnetic field
of the neutron star threads the accretion disk at different radii in a
broad "transition" zone, in spite of predominance of the disk stress
compared with magnetic stress in this region. A schematic diagram of
their model is shown in Figure 2.  Their work is based on the idea of
anomalous resistivity connected with the supposed fast reconnection of
toroidal component of the magnetic field lines across the disk.

Here, we propose an essentially different picture for disk accretion
onto a star with an aligned dipole magnetic field. The idea is based on
the fact that when there is a large difference between the angular
velocity of the star and that of the disk, the magnetic field lines
threading the star and the disk undergo a rapid inflation so that the
field becomes open with separate regions of field lines extending
outward from both the star and the disk. As a result the magnetosphere
consists of an open field line region far from the star and a closed
region approximately corotating close to the star.  This is shown
schematically in Figure 3.  Our model does not assume an anomalous
resistivity, but rather a turbulent magnetic diffusivity of the disk
comparable to the turbulent $\alpha$ viscosity proposed by Shakura (1973).
This
corresponds to a resistivity about $10^4$ times smaller than that assumed
by Ghosh and Lamb (1978,1979a,b).  Campbell (1992) has earlier discussed closed
field line models of disk accretion onto an aligned dipole star assuming
a magnetic diffusivity comparable to the turbulent $\alpha$ viscosity.

Here, we study the matter flow in the disk taking into account the open
magnetic field line region of the magnetosphere. The existence of an
open magnetic field region of the disk leads to the possibility of
magnetically driven outflows.  We analyze the magnetohydrodynamic (MHD)
outflows and their back influence on the disk using the work on MHD
outflows and magnetized disks by Lovelace, Berk, and Contopoulos (1991)
(denoted LBC), Lovelace, Romanova, and Contopoulos (1993) (LRC), and
Lovelace, Romanova, and Newman (1994) (LRN).

In $\S$II we discuss the basic equations.  In $\S$IIA we argue that
part of the star/disk field configuration is open.  In $\S$IIB we give
the basic equations for a magnetized viscous accretion disk.  In
$\S$IIC we discuss magnetically driven outflows and in $\S$IID,
necessary conditon for these outflows.  In $\S$IIE, we give the main
results for the case of outflows where in general star spins up.  In
$\S$IIF, we discuss the magnetic braking of the star by the disk
due to field line twisting, which occurs when
there are no outflows, and show that the star can either spin-down or
spin up depending on the mass accretion rate and the magnetic moment of
the star.  In $\S$III we discuss the conclusions of this work.

\vfill\eject

\nid{\bf{II.~~THEORY}}

The basic equations for an assumed stationary configuration of
plasma are:

$$\nb \cdot (\rho {\bv}) = 0~,~~{\bf{\nb}} \times {\bB} = {4\pi \ov c}
{\bJ}~,~~{\bf{\nb}} \cdot {\bB} = 0~,~~{\bf{\nb}} \times {\bE} = 0~,$$
$${\bJ} = \sm_e (\bE + \bv \times \bB/c)~,~~\rho {\bf{(v \cdot \nb)v}} =
 - \nb
\rm p + \rho {\bf{g}} + {1\ov c} {\bJ} \times {\bB} + {\bF}^{\rm{vis}}~.
\eqno(1)$$  \nid Here, ${\bv}$ is the flow velocity, $\rho$ is the
density, ${\sm_e}$ is the effective electrical conductivity,
$\bF^{\rm{vis}}$ is the viscous force density, $p = \rho k_B T/m$ is the
gas pressure (with $k_B$ the Boltzmann constant, $m$ the mean particle
mass, and with the radiation pressure assumed negligible), and ${\bf{g}}$ is
the
gravitational acceleration.  Outside of the disk, dissipative effects
are considered to be negligible ($\sm_e \rightarrow
\infty,~{\bF}^{\rm{vis}} = 0$, etc.).  We neglect the self-gravity of the disk
and relativistic effects so that ${\bf{g}} = -{\bf{\nb}}
\Phi_g$ with $\Phi_g = - GM/(r^2 + z^2)^{1/2}$, where $M$ is the mass
of the central star.  Equations (1) are supplemented later by a
equation for the conservation of energy in the disk.

A general axisymmetric ${\bB}$-field can be written as ${\bB} = {\bB}_p
+ {\hat{\bf \phi}} B_\phi$, where ${\bB}_p = \nb \times ({\hat{\bf \phi}}
\Psi/r)$ is the
poloidal field, $B_\phi$ is the toroidal field, and $\Psi$ is the flux
function (see, for example, Mestel 1968, or Lovelace et al. 1986).  We use a
non-rotating cylindrical coordinate system so that ${\bB}_p = (B_r, 0,
B_z)$.  Notice that $\Psi (r,z) = const.$ labels a poloidal field line,
$({\bB}_p \cdot {\bnb}) \Psi = 0$, or a flux-surface if the poloidal
field line is rotated about the z-axis.  For the present problem, the
${\bB}_p$ field can be represented as $\Psi = \Psi_* + \Psi'$, with
$\Psi_*$ the star's field-assumed to be a dipole $\Psi_* = \mu r^2 (r^2
+ z^2)^{-{3\ov 2}}$ and with $\Psi'$ due to the non-stellar toroidal
currents.

\nid A.~~ {\bf{Inflation and Opening of Coronal ${\bB}$-Field}}

Figure 4 shows a sketch of the poloidal projections of two nearby field
lines connecting the star and the disk.  Because $\pt /\pt t = 0$, the
${\bE}$ field is
electrostatic and the poloidal plane line integral of ${\bE}$ around the
loop, $1 \rightarrow 2 \rightarrow 3 \rightarrow 4 \rightarrow 1$, is
zero.  Because of axisymmetry and the fact that ${\bE} + {\bv} \times
{\bB}/c = 0$ outside of the disk, the line integrals of ${\bE}$ along
the two curved segments in Figure 2 vanish separately.  That is, the
electrostatic potential is a constant on any given flux surface (see,
for example, Lovelace et al. 1986).  The potential difference between the
points (1,2) on the star's surface is $- {\bf{\dt r}}_{12} \cdot {\bE}_*
= \om_* \dt \Psi/c$, with $\dt \Psi \equiv \Psi_1 - \Psi_2$, where we
have assumed that the star is perfectly conducting $[{\bE}_* = - ({\bv}
\times {\bB})_*/c]$, and where $\om_*$ is the angular rotation rate of
the star.  Thus, we have $\dt r_{34}(E_r)_d = - \om_* \dt \Psi /c = - \om_*
rB_z(r,0)/c$,
where the path element $\dt r_{34}$ is in the mid plane of the disk, and
the "$d\,$" subscript indicates that the quantity is evaluated in the
disk.  In turn, we have $E_r|_d = (J_r/\sm_e - v_\phi B_z/c)_d$.  From
Amp\`ere's law, $J_r = -(c/4\pi) (\pt B_\phi/ \pt z)$, the fact that
$B_\phi$ is necessarily an odd function of $z$, and the approximation
$\pt B_\phi /\pt z = (B_\phi)_h/h$, we have $J_r \approx -
(c/4\pi)(B_\phi)_h/h$, where $h$ is the half-thickness of the disk, and
the notation $(\cdots)_h$ indicates that the quantity is evaluated at $z
= h$.  Combining these results gives for stationary conditions
$$(B_\phi)_h (r) = - {{hr}\ov
\eta_t}[\om_d (r) - \om_*] B_z (r)~,\eqno(2)$$ where $\eta_t = c^2/(4\pi
\sm_e)$ is the magnetic diffusivity of the disk, and $\om_d$ is the
angular rotation rate of the disk.  The fractional variation in $B_z$
between the midplane and surface of the disk is negligible $[O(h/r)]$ if
the disk is thin in the sense that $h/r \ll 1$.  For the conditions of
interest here, we show in $\S$IIB that $h/r \leq c_s/v_K$, where
$c_s \equiv (k_B T/m)^{1\ov
2}$ the Newtonian sound speed based on the mid-plane temperature of the
disk, and $v_K = (GM/r)^{1\ov 2}$ is the Keplerian velocity at $r$.
Campbell (1992) has independently derived equation (2) but does not
place any limit on the ratio $|B_{\phi}/B_z|$ (see discussion following
equation 3).

In contrast with the work of Ghosh and Lamb (1978, 1979a,b), we assume
that the magnetic diffusivity of the disk $\eta_t$ is of the order of
magnitude of the disk's effective viscosity (see Bisnovatyi-Kogan and
Ruzmaikin 1976; Parker 1979; Campbell 1992). Additionally, we assume
that the disk
viscosity is the turbulent viscosity as formulated by Shakura (1973) and
Shakura and Sunyaev (1973), $\nu_t = (2/3) \ap c_s h$, where $\ap$ is a
dimensionless quantity less than unity.  In this work, we assume that
$\ap$ is in the range $10^{-1}$ to $10^{-2}$.  The possible contribution
to the turbulent momentum flux due to small-scale magnetic field
fluctuations is assumed included in $\ap$ (Eardley and Lightman 1975;
Coroniti 1981; Balbus and
Hawley 1992; Kaisig, Tajima, and Lovelace 1992).  We write $\eta_t = D
\nu_t$ where $D$ is dimensionless and of order unity.

With $\eta_t$ the turbulent diffusivity, equation (2) gives
$|(B_\phi)_h /B_z| = (3/2) r |\om_d (r) - \om_*|/(\ap D c_s)$.
This ratio is a measure of the twist of the ${\bB}$ field
between the star and the disk.  For the outer part of the disk
$\om_d(r)$  is expected to be close to the Keplerian value
$\om_K = (GM/r^3)^{1\ov 2}$.  Thus, it would appear that the
twist $|(B_\phi)_h/ B_z|$ can be very much larger than unity,
say, $> 10$.  We argue here that such large values of the
twist do not occur.

For a discussion of the limitation on the twist $|(B_\phi)_h /B_z|$ we
consider that the plasma outside of the disk is force-free, ${\bJ}
\times {\bB} = 0$, which is the coronal plasma limit of Gold and Hoyle
(1960).  This limit is valid under conditions where the kinetic energy
density of the plasma is much less than ${\bB}^2/8\pi$.  The general
response of force-free coronal magnetic field loops to stress
(differential twisting) applied to the loop foot points (in the solar
photosphere) has been studied intensely by Aly (1984, 1991), Sturrock
(1991), and Porter, Klimchuk, and Sturrock (1992).  The conclusion of
these studies is that a closed field loop with a small twist evolves
into an open field-line configuration as the twist is increased.  The
related problem of the twisting of a force free magnetic field
configuration with foot points at different radii in a Keplerian disk
has been studied by Newman, Newman and Lovelace (1992), and Lynden-Bell
and Boily (1994).  The twisting due to
the differential rotation of the disk acts to increase the total
magnetic energy of the coronal field and this in turn acts to ``inflate"
the field as shown in Figure 5.

An alternative measure of the field twist between the star and the disk
is simply the difference in the azimuthal location, $\Dt \phi$, of the
stellar and disk foot points of a given flux tube.  We have $r
d\phi/dl_p =  B_\phi (r,z)/|{\bB}_p(r,z)|$, where $dl_p$ is the length
element along a poloidal field line $\Psi (r,z) = const.$  Because $r
B_\phi (r,z)$ depends only on $\Psi$ in a force-free plasma, we have
$$\Dt \phi = [r (B_\phi)_h]_{disk} \int_{star}^{disk} {{dl_p}\ov
{r|{\bf{\nb}} \Psi|}}~. \eqno(3)$$ \nid For small $\Dt \phi$ (say, $<1$),
the poloidal field is close to that of a dipole, and the integral
gives $\Dt \phi = C_t [(B_\phi)_h /B_z]_d$ with $C_t=8/15$.  For
increasing $\Dt \phi$, the value of $C_t$ increases because of the
longer path length and the smaller value of $|{\bf{\nb}} \Psi|$.
The present  situation is
analogous to a current-carrying plasma column with a longitudinal $B$
field.  For the plasma column there is the well-known Kruskal-Shafranov
stability condition (see, for example, Bateman 1980) on the twist $\Dt
\phi$ of the field (at the column's surface) over the length of the
column.  This condition, $\Dt \phi < 2\pi$, is required for stability
against symmetry changing kink perturbations.  A related,
non-axisymmetric instability may accompany the above mentioned inflation
and opening of the star/disk field in response to increasing $\Dt \phi$.

We propose that there is a definite upper limit on the twist $\Dt \phi$
of any field lines connecting the star and the disk.  This limit implies
a corresponding limit on $|(B_\phi)_h /B_z|_d$ in equation (2).  If $\Dt
\phi$ is larger than this limit, then the field is assumed to be open as
indicated in Figure (3). Note that when the field threading the disk is open
equation (2) does not apply.

\nid {\bf{B.~~Basic Equations for the Disk}}

The flow velocity in the disk is ${\bv}(r)=-u(r)\hat
r+v_\phi(r)\hat\phi$, where $u$ is the accretion speed.  The  surface
density is $\sigma=\int\limits_{-h}^h dz\,\rho(r,z)$. The main disk equations
are:

\medskip
\nid 1.~~~Mass conservation:
$$ \dot M=2\pi r\sigma u,~=~const. \eqno(4)$$

\item{} We assume that the change of $\dot M$ with $r$ due to possible MHD
outflows is small.

\nid 2.~~~Radial force balance:
$${\sigma v_\phi^2\over r}={GM\sigma\over r^2}-{(B_rB_z)_h\over 2\pi}
+{d\over dr}\int\limits_{-h}^h dz\,p~~. \eqno(5)$$

\nid 3.~~~Angular momentum conservation:
$$\eqalignno{ {d\ov dr}(\dot M F) &= -r^2 (B_\phi B_z)_h, \cr
F &\equiv r^2 \om + {{r^2 \nu_t}\ov u} {{d\om}\ov dr}~,&(6)}$$

\item{} where $\om = v_\phi/r$.  The term $-r^2 (B_\phi B_z)_h$
represents the outflow of angular momentum from the $\pm z$ surfaces of
the disk (that is, a torque on the disk).  We assume that the angular
momentum carried by the matter of the outflow is small compared with
that carried by the field.  If the $B$ field at $r$ is open as discussed
in $\S$IIA, then the angular momentum outflow from the disk is carried
to infinity.  On the other hand, if the field is closed, then the
angular momentum outflow from (or inflow to) the disk is carried by the
coronal $B$ field to (or from) the star.

\nid 4.~~~Energy conservation:
$$\sigma\nu_t\left(r{d\omega\over dr}\right)^2 +{4\pi\over
{c^2}}\int\limits_{-h}^h dz
{}~{\eta_t{\bJ}^2} ={{4acT^4}\over {3 \kp \sigma}}\equiv 2 \sigma_B T_{eff}^4~.
\eqno(7)$$

\item{} The first term is the viscous dissipation, the second is the Ohmic
dissipation, while the third term is the power per unit area carried
off by radiation (in the $\pm z$ directions) from the disk which is
assumed optically thick.  Here, $\kappa$ is the
opacity assumed due to electron scattering, $a$ and $\sigma_B = ac/4$ are
the usual
radiation constants,  $T$ is the midplane temperature of the disk, and
$T_{eff}$
is disk's effective surface temperature.

\nid 5.~~~Conservation of magnetic flux for a general time-dependent
disk:
$${\pt \ov \pt t}(rB_z)={\pt \ov \pt r}\left[rB_zu-{\eta_t r\over
h}(B_r)_h\right], \eqno(8a)$$
\item{} where the generally small radial diffusion of the $B_z$ field
has been neglected (see LRN).  The first term inside the square brackets
represents
the advection of the field while the second term represents the
diffusive drift.  In a stationary state,
$$\beta_r (r) \equiv {{(B_r)_h}\ov
B_z} ={ {uh}\ov \eta_t}.\eqno(8b)$$

\nid 6.~~~Vertical hydrostatic equilibrium:
\item{} The $z$-component of the Navier-Stokes
equation (1) gives the condition
for vertical hydrostatic balance which can be written as $$
\left({h\over r}\right)^2 + b \left({h\over r}\right) - ({ c_s\over
v_K})^2 = 0~,\eqno(9)$$ where $c_s$ is the Newtonian sound speed
based on the midplane temperature of the disk,
and $b \equiv r\left\lbrack \left(B_r\right)_h^2 +
\left(B_\phi\right)_h^2\right\rbrack / \left(4\pi\sigma v_K^2\right)$
(Wang, Lovelace, and Sulkanen 1990).  Radiation pressure is
negligible for the conditions of interest.  For $b \ll 2c_s/v_K$,
this equation gives the well-known relation $h/r = c_s/v_K
$ (Shakura and Sunyaev 1973),  while for $b\gg 2c_s/v_K$
it gives $h/r = b^{-1}(c_s/v_K)^2$
which is smaller than $c_s/v_K$ owing to the
compressive effect of the magnetic field external to the disk (Wang et al.
1990)
It is useful to write $b = \epsilon~ \beta^2$,
where $\epsilon \equiv r B_z^2(r,0)/(2\pi\sigma v_K^2)$ and $\beta^2 \equiv
[(B_r)_h^2+(B_\phi)_h^2]/(2B_z^2)$.  In $\S$IID, we discuss
that $\beta = O(1)$ in order to have outflows from the disk, and $\beta \leq 1$
with no outflows.  Thus, the Alfv{\'e}n speed in the midplane of the disk is
$v_A = v_K\epsilon^{1\over2}(h/r)^{1\over2}$.  We may term the magnetic
field as weak if $\epsilon < c_s/v_K \approx h/r$.  In this limit,
$v_A/c_s \approx (\epsilon r/h)^{1\over2} \leq 1$, and the
magnetic compression of the disk is always small.  In the opposite, strong
field limit, $\epsilon > c_s/v_K$.  If at the same time $\beta = O(1)$, then
the disk is magnetically compressed, and for $\epsilon \gg c_s/v_K$, $v_A/c_s
= 1/\beta$.
\ms

The following subsections involve applications of equations (4) -- (8).
Note that with the magnetic field terms neglected these equations give
the Shakura-Sunyaev (1973) solution for region ``$b$".

\nid{\bf{C.~~Magnetically Driven Outflows}}

Consider the outer region of the disk where the value of the field twist
$\Dt \phi$ given by equation (3) [or $|(B_\phi)_h/B_z|$ of equation (2)]
is larger than a critical value.  In this region the $B$ field threading
the disk is open as discussed in $\S$IIA.  The angular momentum flux
carried by the disk is $\dot M F$.  If the angular rotation of the disk
is approximately Keplerian, $\om (r) \approx \om_K (r)$, then the
viscous transport contribution to $F$ is $r^2 (\nu/u)(d\om/dr) \approx -
(3/2) r^2 \om [h/(\beta_r r)]$.  As discussed in detail below a necessary
condition for MHD outflows is that $\beta_r = O(1)$.  Consequently, the
viscous contribution to $F$ in equation (6) is smaller than the bulk
transport term by
the small factor $h/r << 1$, and $F \approx \om r^2$.

In equation (6) we therefore have $d(\dot M F)/dr \approx \dot M \om_K
r/2$.  In equation (6) we let $(B_\phi)_h = -\beta_\phi B_z(r,0)$.  Studies of
MHD outflows (Blandford and Payne 1982, LBC, LRC)
indicate that $\beta_\phi = const. \los O(1)$.  As a result, equation (6)
implies that
$$[B_z(r,0)]_w = k /r^{5/4},~ {\rm{with}}~ k = [\dot M (GM)^{1\ov 2} /(2
\beta_\phi)]^{1\ov 2}~,\eqno(10)$$
\nid where the $w$ subscript indicates the wind region of the disk.
Of course at a sufficiently small radius, denoted $r_{wi}\;$, the outflow
ceases, and the dependence of $B_z(r,0)$ reverts to approximately the
stellar dipole field, $\mu/r^3$.  As a consistency condition we must
have
$${\cE} = (k /r_{wi}^{5/4}){\Big /}(\mu /r_{wi}^3) = \bigg [ {{\dot M
(GM)^{1/2}r_{wi}^{7/2}}\ov {\mu^2 (2\beta_\phi)}}\bigg ]^{1\ov 2} < 1~.
\eqno (11)$$

\nid The transition between the dipole and outflow field dependences is
handled by letting $B_z (r,0) = (\mu /r^3) g(r) + (k /r^{5/4}) [1 -
g(r)]$, where, for example, $g(r) = \{ 1 + \exp [(r - r_{wi}) / \Dt r ]
\}^{-1}$ and $\Dt r/r_{wi} \ll 1$.

Equations (4) -- (9) can be used to obtain the disk parameters in the
region of outflow.  As shown below the viscous dissipation is much
larger than the Ohmic, and equation (7) then gives $T = [81 \dot M^2
(GM)^{1/2}\kp \big /(128\pi^2 \ap D \beta_r^2 ac)]^{1\ov 4} r^{-{7\ov 8}}$.
 For illustrative values, $M = 1M_\odot$, $\dot M = 10^{17}$g/s $\approx
1.6 \times 10^{-9} M_\odot$/yr, $\ap = 0.1$, $\beta_r = 0.58$, and $D =
1$, we find at the representative distance $r = 10^8$cm that $T \approx
0.44 \times 10^6K$, $v_K = 1.15 \times 10^9$cm/s, $c_s = 7.6 \times
10^6$cm/s, $u = 0.88 \times 10^6$cm/s, $\sm = 180$g/cm$^2$, and $h/r =
6.6 \times 10^{-3}$.  The ratio of the ohmic dissipation to that due to
viscosity is: $(4/27) \ap D \beta_r^3 \ll 1$.  The ratio of the viscous
dissipation to the power output of the outflows (the $\pm\, z$ Poynting
fluxes)
(per unit area of the disk) is: $(9/2)(h/r)(\beta_r D)^{-1}
\ll 1$.  Thus, most of the accretion power for $r > r_{wi}$
goes into the outflows.  However, the fraction of the total accretion
power in outflows is small if $r_{wi} \gg r_*$, where $r_*$ is the
star's radius ($\sim 10^6$ cm for a neutron star).  Note that $\dot M =
10^{17}$g/s and $M = 1 M_\odot$ correspond to a total accretion power or
luminosity $L_0 = GM\dot M/r_* \approx 1.33 \times
10^{37}$(erg/s)($10^6$cm/$r_*$).

Conservation of the poloidal flux threading the disk implies that there
is an outer radius, denoted $r_{wo}$, of the outflow from the disk.  That
is, we have $\int_{r_{wi}}^{r_{wo}} rdr (k /r^{5/4}) = \int_{r_{wi}}^{r_{wo}}
rdr (\mu /r^3)$.  Thus, the outer radius is $r_{wo} \approx
r_{wi} (f {\cE})^{-{4/3}}$, where $f \approx 1$ for $1-{\cE} \ll 1$
and $f =3/4$ for ${\cE} \ll 1$.  For $r > r_{wo}$, $B_z(r,0)$ rapidly
approaches zero.

Equation (9) gives the disk thickness.  From equation (10),
we have $\epsilon = rB_z^2/(2\pi\sigma v_K^2)$ $ =
(u/c_s)(c_s/v_K)/(2\beta_\phi)$.
The magnetic compression of the disk is small if $\epsilon < c_s/v_K$ or
equivalently if $u/c_s < 1$.  Our numerical solutions have $u/c_s < 1$ in the
region of outflow.

\nid {\bf{D.~~ Necessary Condition for MHD Outflows}}

A necessary condition on $\beta_r$ for MHD outflows from the disk can be
obtained by considering
the net force ${\cF}_p$ on a fluid particle in the
direction of its poloidal motion above the disk.
Notice that for stationary flows, the poloidal
flow velocity ${\bv}_p$
is parallel to ${\bB}_p$ owing to the assumed axisymmetry and perfect
conductivity.  However, in general ${\bv} \times {\bB} \not= 0$ so that
the fluid particles do not move like ``beads on a wire".  Above, but close
to the disk ($h \leq z \ll r$), we assume that the poloidal field lines
are approximately straight.  Thus, the poloidal position of a fluid
particle above the disk is ${\bf r}=(r_o+Ssin(\theta)){\hat {\bf r}} +
Scos(\theta){\hat {\bf z}}$,
where $S$ is the distance along the path from the starting value $S_o =
h/cos(\theta)$ at a radius $r_o$, $tan(\theta)=(B_r)_h/B_z=\beta_r$, and
$z=Scos(\theta)$.  The effective potential is
$U(S)=-(1/2)\om_o^2(r_o+Ssin(\theta))^2-GM/|\bf r|$~(LRN),
and the force in the direction of the particle's poloidal motion
is ${\cF}_p =-\partial U/\partial S$.  For distances $z$ much less
than the distance to the Alfv{\'e}n point, $\om_o \approx \om_K(r_o)$.
In this way we find
$${\cF}_p = - (\om_K^2 - \om^2)r \biggr ({{B_r} \ov |{\bB}_p|}\biggr ) +
\om_K^2 z\; (3 \beta_r^2 - 1) \biggr ({{B_z} \ov |{\bB}_p|}\biggr )~,
\eqno(12)$$
\nid for $h \los z \ll r$, where we
assume $B_z > 0$.

The slow
magnetosonic point of the outflow occurs where ${\cF}_p = 0$ (LRC) at
the distance
$$z_s = r \biggr ( 1 - {\om^2 \ov \om_K^2} \biggr )  {\beta_r \ov
{3\beta_r^2 -1}}~. \eqno (13)$$

\nid The factor $1 - (\om/\om_K)^2$ can be obtained from the radial force
balance equation (5).  The radial pressure force is small compared with
the magnetic force for the conditions we consider, $\ap
\beta_r^2/(3\beta_\phi) > h/r \ll 1$.  Thus,
$${{z_s} \ov h} = \Big({{2 \ap DB_z^2 r}\ov {3 \dot M \om_K} }\Big)
{\beta_r^3\ov {3 \beta_r^2 - 1}}. \eqno(14)$$
\nid For $r_{wi} < r < r_{wo}$, the quantity in brackets is simply $\ap
/ (3 \beta_\phi)$ owing to equation (10).  Note that the minimum of
$\beta_r^3/(3\beta_r^2 - 1)$ is $1/2$ at $\beta_r = 1$.  If the gas
near the surface of the disk ($h \los z \leq z_s$) is assumed isothermal
with temperature $T_1 \ll T$ and sound speed $c_{s1} \equiv (k_B
T_1/m)^{1\ov2}$,
then the density at the slow magnetosonic point $\rho_s$ can be obtained
using the MHD form of Bernoulli's equation (LRC).  At the slow magnetosonic
point, the poloidal flow speed is $v_p = c_{s1}|{\bB}_p|/|{\bB}| \equiv
v_{sm}$,
the slow magnetosonic speed, and $\rho_s = \rho_h exp\left\{-1/2 -[U(z_s)
-U(h)]
/c_{s1}^2\right \}$, where U is now regarded as a function of $z$.  In turn,
the mass flux density from the $+z$ surface of the disk is $\rho_s v_{sm}
cos(\theta) = \rho_s c_{s1} B_z/|{\bB}|$.

The consistency of the magnetically driven outflow solutions requires
$z_s \gos h$.  If this were not the case, the outflows would be matter
rather than field dominated near the disk.  There are two possible
regimes having $z_s \gos h$: one is with $\ap D/(3\beta_\phi) < 1$ [or
$\ap D/(3 \beta_\phi) \ll 1$] and $\beta_r \gos 1/\sqrt 3$; another is with
$\ap D/(6 \beta_\phi) > 1$ and $\beta_r > 1$.  In the present work, we
consider the first regime which is consistent in the following respect:
The magnetic pinching force on the inner part of the outflow $(r <
r_{wo})$ increases as the mass flux in the outflow increases (LBC).  In
turn, an increase in the pinching force acts to decrease $\beta_r =
(B_r)_h/B_z$ thereby increasing $z_s/h$ and decreasing the mass flux in
the outflow.

In addition to the condition $z_s \gos h$, the slow magnetosonic point
must be at a distance $z_s$ {\it{not}} much larger than $h$, say $2h$,
in order for the outflow to have a non-zero mass flux.  At the inner
radius of the outflow, $r_{wi}\;$, the $B_z(r,0)$ field in the disk is
$\mu/(2 r^3_{wi})$ as discussed in $\S$IIC.  We then have
$$r_{wi} \gos \bigg [ {{\ap D \mu^2}\ov {24 \dot M(GM)^{1\ov 2}}}\bigg ]^{2\ov
7} \approx 0.75 \times 10^8{\rm{cm}} \bigg [ \Big( {\ap D \ov 0.1}\Big)
\Big ( {\mu \ov {10^{30} {\rm{Gcm}}^3} } \Big )^2 \Big
({{10^{17}{\rm{g/s}}}\ov \dot M}\Big) \Big( {M_\odot \ov M}\Big )^{1\ov
2}\bigg ]^{2\ov 7}~,\eqno (15)$$
\nid from equation (14).  Note that from equation (11), we get ${\cE}
\gos [\ap D/(48 \beta_\phi)]^{1\ov 2}$ or ${\cE} \gos 0.046$ and $r_{wo}
\los 90 r_{wi}\;$ for $\ap = 0.1$, $D = 1$ and $\beta_\phi = 1$.

\nid{\bf{E.~~MHD Outflows and Spin-Up ($r_{to} < r_{cr}$)}}

We have numerically integrated equations (4) -- (9) assuming
magnetically driven outflows from the disk as discussed in $\S\S$IIC \&
D.  We integrate the equations inward starting from a large radial
distance $r < r_{wo}\;$.  The inner radius of the region of outflow is
determined as indicated in $\S$IID.  For $r < r_{wi}\;$, $d F/dr = 0$,
and the solutions in this range of $r$ all exhibit a ``turn-over
radius", $r_{to}\;$, where $d\om /dr =0$.  This ``turn over" results
from the radially outward magnetic force in equation (5) which becomes
stronger as $r$ decreases.  In the region close to the turnover, the
magnetic compression of the disk becomes strong in that $\epsilon = O(1)$.
Figure 6 shows a representative solution.
{}From a least squares fitting of many integrations, we find
$$r_{to} \approx 0.91 \times 10^8{\rm{cm}} \Big ( {\ap D \ov 0.1}\Big
)^{0.3} \Big ( {\mu \ov {10^{30}{\rm{G cm}}^3} }\Big)^{0.57} \Big (
{{10^{17}{\rm{g/s}}}\ov \dot M}\Big)^{0.3} \Big ( {M_\odot \ov
M}\Big)^{0.15}~. \eqno (16)$$
Figure 7 shows as an example the dependence of $r_{to}$ on $\mu$.  The
turn-over radius is less than but in all cases close to $r_{wi}$.  Thus,
equation (16) is compatible with equation (15).  Our radius $r_{to}$ has
a role similar to that of the Alfv\'en radius $r_A$ of Ghosh and Lamb
(1978, 1979a,b).  The dependences we find of $r_{to}$ on $\mu$, $\dot
M$, and $M$ are close to those of $r_A$.  However, our analysis shows an
important dependence on $\ap D$ which is proportional to the magnetic
diffusivity of the disk.  Furthermore, our $r_{to}$ is smaller
than $r_A$ by a factor of about $(\ap D/12)^{0.3}$, which is $\approx
0.24$ for $\ap D = 0.1$.

The turn-over radius is important in the respect that the inward angular
momentum flux carried by the disk $(\dot M F)$ is $\dot M \om_{to}
r_{to}^2$ because $d\om/dr = 0$ at $r = r_{to}$.  [Note that we do not
consider here the possibility of significant poloidal current flow along
the open field lines extending from the polar caps of the star ($B_\phi
> 0$ for $z > 0$ in figure 3) which would act to remove angular momentum
from the star to infinity.]  From numerical integrations, we find to
a good approximation that $\om_{to}
= (GM/r^3_{to})^{1\ov 2}$, so that the influx of angular momentum to the
star is $\dot M F_{to} = \dot M (GM r_{to})^{1\ov 2}$.  Thus, the rate
of increase of the star's angular momentum $J$ is:
$${{dJ}\ov {dt}} = \dot M (GM r_{to})^{1\ov 2}~. \eqno (17)$$
\nid With $I$ the moment of inertia of the star, $J = I \om_*$,
and $dJ/dt = (dI/dM)\dot M\om_* + I (d\om_*/dt)$.  For the situation
of interest, the term proportional to $dI/dM$ is negligible because
$r_{to}$ is much larger than the star's radius.
Thus, we have ``spin-up" of the star,
$${{d\om_*} \ov {dt}} = {{\dot M (GM r_{to})^{1\ov 2}}\ov I}~,
\eqno(18a)$$
\nid or
$${1\ov P} {dP \ov dt} \approx -5.8 \times 10^{-5} {1\ov {\rm{yr}}}
\Big( {P\ov 1s} \Big) \Big( { \dot M\ov {10^{17}}{\rm{g/s}} }\Big) \Big(
{ {10^{45}{\rm{g cm}}^2}\ov I } \Big) \Big( {M\ov M_\odot} \Big)^{1\ov
2} \Big( { r_{to} \ov {10^8{\rm{cm}}} } \Big)^{1\ov 2}~,\eqno(18b)$$
\nid  where $P
= 2\pi/\om_*$ is the pulsar period.  Because the total accretion luminosity
$L=GM{\dot M}/r_*$ with $r_*$ the star's radius, ${\dot P}\propto -P^2 L^{0.85}
\mu^{0.285}$ for constant $\alpha D, r_*,$ and $ M$.

Our outflow solutions are consistent only in the case where $\om_{to} =
(GM/r_{to}^3)^{1\ov 2} > \om_*$.  In this case, with $r$ decreasing from
$r_{to}$, the rotation rate of the disk $\om(r)$ decreases and
approaches the rotation rate of the magnetosphere $(\approx \om_*)$ from
above where it matches onto the magnetospheric rotation through a
radially thin turbulent boundary layer.  We find no consistent stationary
solutions when $\om(r)$ decreases below $\om_*$.  The condition
$\om_{to} > \om_*$ is the same as
$$r_{to} < r_{cr} \equiv \bigg ( {{GM}\ov \om_*^2}\bigg )^{1\ov 3} \approx 1.5
\times 10^8{\rm{cm}}\bigg ({M \ov M_\odot}\bigg)^{1\ov 3} \bigg ({P \ov
{1s}}\bigg)^{2\ov 3}~,\eqno (19)$$
\nid where $r_{cr}$ is commonly referred to as the ``co-rotation
radius".  Note that $r_{cr}$ decreases during spin-up.  Thus,

$r_{cr}$ approaches $r_{to}$ if $r_{to}$ does not decrease rapidly.

\nid{\bf{F.~~Magnetic Braking of Star by Disk; Spin-Up or Spin-Down ($r_{to} >
r_{cr}$)}}

For $r_{to} > r_{cr}$, the $B$ field in the outer part of the disk $(r
\gg r_{cr})$ is open as discussed in $\S$IIA, but there are no
consistent stationary  MHD
outflows, $(B_\phi)_h = 0,~\beta_r = h u/\eta_t \ll 1$, and the Ohmic
dissipation is small compared with that due to viscosity.  Thus, the
outer part of the disk obeys essentially the equations of Shakura and
Sunyaev (1973) for a Keplerian disk with $F = const. \equiv F_\infty$ in
equation (6) undetermined.  We show below that $F_\infty$ can be
determined self-consistently by considering the region near $r_{cr}$ where
the Keplerian accretion flow is
brought into co-rotation with the star.  Because there are no
outflows, the angular momentum influx to the star is $\dot M F_\infty$
which is equal to $d J_*/dt$.

In the region of the disk where $\om (r)$ is close to $\om_*$, that is,
where $r \sim r_{cr}$, $|(B_\phi)_h/B_z|$ given by equation (2) is
{\it{not}}  much larger than unity.  In this region, closed but
twisted field lines link the star and the disk.  The twist of the field
acts to remove (or add) angular momentum from (to) the disk if $B_\phi
B_z > 0$ (or $< 0$).  The angular momentum removed from the disk is
deposited on the star via the $B$ field.  In this region of the disk the
key equations are (2) and (6) which we rewrite as
$$\eqalignno{ {d\om \ov dr} &= {u \ov {r^2\nu_t}} (F - \om
r^2)~,&(20a)\cr
{dF \ov dr} &= {hr\ov D\nu_t}(\om - \om_*)\Big( {{r^2 B_z^2}\ov \dot
M}\Big) H \Big( {\tau\ov\tau_{max}}\Big)~.&(20b)\cr}$$
\nid Here, $H(x)$ is a Heaviside function such that $H(x) = 1$ for $|x|
< 1$ and $H(x) = 0$ for $|x| > 1$;
$$\tau(r) \equiv { {(B_\phi)_h (r)} \ov {B_z(r,0)} } = -{hr \ov {D
\nu_t}} (\om -\om_*)\eqno(20c)$$
\nid is a measure of the field twist;  $\tau_{max} = const.$ is the
maximum value of the twist;  and
$$F(r) \equiv r^2 \om + {{r^2 \nu_t}\ov u}{{d \om}\ov dr}~.\eqno(20d)$$
\nid We consider $\tau_{max}$ to be a universal
constant.  For $|\tau| > \tau_{max}$ the field configuration becomes
open.  In equations (18), $B_z$ is assumed to be the star's
dipole field.  The other equations of $\S$IIB are still needed.  Note
that both the viscous and Ohmic heating must be retained in the energy
equation (7). The viscous dissipation is dominant for $r > r_{cr}$,
while the Ohmic dissipation dominates for $r < r_{cr}$.  Equations (5)
and (8b) can be combined to give $u$ as a function of $r, T,$ and $\omega$.
In turn, this expression for $u$ can be combined with the energy conservation
equation (7) to give both $u$ and $T$ as functions of $r, \omega,$ and
$\partial
\omega/\partial r$ if viscous heating dominates, or as functions of $r$ and
$\omega$ if Ohmic heating dominates.  Furthermore, $\nu_t = (2/3)\alpha c_s h$
in equation (20) can be derived from $T$ and $h/r$ from equation (9) using
$\beta_r$ from equation (8b), $\beta_\phi = - \tau$, and $\epsilon$ from
$B_z$ and $u$.

We have solved the equations (5, 7-9, \& 20) by numerical integration
starting from the outside, at a distance where $\tau >
\tau_{max}$ and $F = const. = F_\infty$, and integrating inward through
$r_{cr}$.  The fact that the value of $F_\infty$ is not known a priori
points to the use of a ``shooting method" for its determination.  Using
this approach, we find that there is in general a unique value of
$F_\infty$, denoted $F_\infty^0$, such that the solution for $\om(r)$
smoothly approaches $\om_*$ for $r$ decreasing below $r_{cr}$.  If
$F_\infty$ is smaller than $F_\infty^0$, then $\om(r)$ follows the
Keplerian law as $r$ decreases below $r_{cr}$.  On the other hand, if
$F_\infty$ is larger than $F_\infty^0$, then $\om(r)$ goes through a
maximum and decreases rapidly to values much less than $\om_*$.  The
only physical solution is that with $F_\infty = F_\infty^0$.
Hereafter, the zero superscript on $F_\infty$ is implicit.

Figure 8 shows the radial dependences of the main physical quantities
for a case of magnetic braking.
The behavior of equations (20) can be
understood qualitatively by noting that if $\om(r)$ were to decrease
linearly through $r_{cr}$, then the radial width of the region of
braking, $dF/dr < 0$ where $\om < \om_*$, would equal that the width of
the region of acceleration, $dF/dr > 0$ where $\om > \om_*$.
Consequently, $F$ would change by only a small fractional amount.
However, if within the braking region $F(r)$ attains a value equal to
$\om r^2$, then $d\om/dr = 0$ at this point and $d\om /dr$ remains small
and $\om \approx \om_*$ for smaller $r$. A rough estimation gives the
radial width of the braking region as $\Dt r \sim \tau_{max} Dc_s/\om_K$,
where $c_s$ is based on the temperature just outside of this region,
and the jump in $F$ as $\Dt F \sim \tau_{max} \Dt r (r^2B_z^2/\dot M)$.

Figure 9 shows the dependence
of $F_\infty$ on $\mu$ and on $\dot M$ for sequences of cases where
$r_{cr}$ is assumed comparable to $r_{to}$.  This is clearly not a
necessary choice; that is, $r_{to}$ could differ significantly from
$r_{cr}$.  However, with $r_{cr} = r_{to}$, Figure 9
allows a direct comparison of the angular momentum influx to the star
($\dot M F_K$) for the case of outflows $(r_{to} \los r_{cr})$ with the
influx $(\dot M F_\infty)$ in the case of no outflows and magnetic
braking $(r_{to} > r_{cr})$.  Figure 9 suggests that a bimodal behavior
can occur where the star switches between spin-up and spin-down, owing
to some extraneous perturbation.

\vfill\eject

\nid{\bf{III.~~DISCUSSION}}

This work has made a new investigation of the problem of disk accretion
of matter onto a rotating star with an aligned dipole magnetic field.
This is a significant idealization in that the relevant objects are
mis-aligned X-ray pulsars in binary systems.  Nevertheless, the aligned
case is important for understanding the theory and as guide to theory of
the mis-aligned rotator.

The important new ideas introduced in this work are: (1) When the angular
velocity of the star and disk differ substantially, the $B$ field
between the star and disk rapidly ``inflates" (see Figure 5) and this
gives rise to regions of open field lines extending from the polar caps
of the star to infinity and from the disk to infinity as shown in Figure
3.  (As in earlier models, there is a closed donut shaped
magnetosphere surrounding the star where the matter rotates at about the
angular rate of the star.) (2) The open field line region of the disk
may have magnetically driven outflows or winds from its $\pm z$
surfaces, and these outflows can have an important back reaction on the
disk structure.  We find that outflows occur when the ``turn over
radius" $r_{to}$ of equation (16) is less than the ``co-rotation radius"
$r_{cr} = (GM/\om_*^2)^{1\ov 3}$, where $\om_*$ is the rotation rate of
the star.  The solutions we find with outflows from the disk give an
influx of angular momentum to the star, that is, the star spins up.
Thus, a system initially with $r_{to} < r_{cr}$ evolves with $r_{cr}$
decreasing towards $r_{to}$ if $r_{to}$ does not decrease rapidly.  A
necessary condition for MHD outflows is that the slow magnetosonic
point of the outflow be located relatively close to the disk surface,
which is possible if the angular rotation rate of the disk $\om$ is
less than the Keplerian rate by a small fractional amount and $\beta_r =
(B_r)_h/B_z = O(1)$.  If $\om$ is signifcantly less than the Keplerian
value, MHD outflows do not occur.  Under these conditions the angular
momentum flux in the disk is carried mainly by the bulk transport of
the accreting matter and it is inward; that is, the viscous transport
is small.

In our treatment of the disk, we assume the ``$\ap$" turbulent viscosity
model of Shakura and Sunyaev (1973), and, further, that the magnetic
diffusivity of the disk is of the order of the turbulent viscosity
(Bisnovatyi-Kogan and Ruzmaikin 1976; Parker 1979).  This magnetic
diffusivity is smaller than the anomalous diffusivity invoked by Ghosh
and Lamb (1978, 1979a, b) by a factor of order $10^{4}$. Also, in
contrast with the model of Ghosh and Lamb, accreting matter inside
of $r_{to}$ or $r_{cr}$ (if $r_{cr}<r_{to}$) remains in a disk.
Funnelling of matter onto the star's magnetic poles may result from
from mis-alignment of the stellar dipole field.

When $r_{to} > r_{cr}$, we argue that outflows do not occur but rather
that there is ``magnetic braking" of the accretion flow in the vicinity
of $r_{cr}$.  The braking solutions we find can give spin-up or
spin-down (or no spin change) of the star depending mainly on the star's
magnetic moment and the mass accretion rate. In contrast with the
case of outflows, the angular momentum flux in the disk is due to
both the bulk transport (inward) and the viscous transport (outward).
For braking solutions with
$r_{cr}$ comparable with $r_{to}$, we find spin-down of the star for
large values of the star's magnetic moment and/or large values of the
mass accretion rate.  Thus, for a system with $r_{cr}$ comparable to
$r_{to}$, bimodal behavior is possible where the system flips between
spin-up $(r_{to} < r_{cr})$ and spin-down.  This transistion could be
triggered by a number of factors, for example, intermittency of $\ap$,
$B$-field flux introduced from the companion star, or variation in the
mass accretion rate, owing to the dependence of $r_{to}$ on $\ap,~\mu$,
and $\dot M$.  The pulsar $GX~1 + 4$ shown in Figure 1 illustrates the
bimodal behavior observed for some objects.  Future work is planned to
address the time-dependent accretion onto an aligned dipole rotator
including magnetically driven outflows from the disk.

\nid {\bf{ACKNOWLEDGEMENTS}}

One of us (M.M.R.) thanks Professor Y. Terzian for hospitality of the
Department of Astronomy at Cornell where part of this work was done.  We
thank Drs.
L. Bildsten, C. Campbell, C. Clarke, L. Mestel, R. Pudritz, and I. Wasserman
for stimulating
discussions and valuable comments on this work.  We thank Ms.
Xuan Campbell for preparation of the manuscript.  The work of
M.M.R. and G.B-K. was supported by Russian Fundamental Research Foundation
Grant No
93-02-17106. The work of M.M.R. was also supported
by the Scientific and Educational Center of
Kosmomicrophysics ``KOSMION". The work of R.V.E.L. was supported in part
by NASA grant NAGW 2293.  The authors were also supported in part by NSF grant
AST-9320068.

\vfill\eject
\baselineskip=12pt

\centerline{\bf{ REFERENCES}}
\bs
\item{} Aly, J.J. 1984, ApJ, 283, 349
\ms
\item{} Aly, J.J. 1991, ApJ (Letters), 375, L61
\ms
\item{} Balbus, S.A., \& Hawley, J.F. 1992, ApJ, 400, 610
\ms
\item{} Bateman, G. 1980, MHD Instabilities (Cambridge: MIT Press) Chap. 6
\ms
\item{} Bildsten, L. 1993, private communication
\ms
\item{} Bisnovatyi-Kogan, G.S., \& Friedman, A.M., 1970, Sov.
Astron., 13, 177
\ms
\item{} Bisnovatyi-Kogan, G.S., Goncharskii, A.V., Komberg, B.V.,
Cherepashchuk, A.M. \& Yagola, A.G. 1977, Sov. Astron., 21, 133
\ms
\item{} Bisnovatyi-Kogan, G.S., \& Ruzmaikin. A.A. 1976, Astroph. and
Spa. Sci, 42, 401
\ms
\item{} Blandford, R.D., \& Payne, D.G. 1982, MNRAS, 199, 883
\ms
\item{} Campbell, C.G. 1992, Geophys. Astrophys. Fluid Dynamics, 63,179
\ms
\item{} Coroniti, F.V. 1981, ApJ, 244, 587
\ms
\item{} Davidson, K., \& Ostriker, J.P. 1973, ApJ, 179, 585
\ms
\item{} Davies, R.E., \& Pringle, J.E. 1981, MNRAS, 196, 209
\ms
\item{} Eardley, D. M., \& Lightman, A. P. 1975, ApJ, 200, 187
\ms
\item{} Fabbiano, G., \& Schreier, E.J. 1977, ApJ, 214, 235
\ms
\item{} Ghosh, P., \& Lamb, F.K. 1978, ApJ (Letters),223, L83
\ms
\item{} Ghosh, P., \& Lamb, F.K. 1979a, ApJ, 232, 259
\ms
\item{} Ghosh, P., \& Lamb, F.K. 1979b, ApJ, 234, 296
\ms
\item{} Giacconi, R., Gursky, H., Kellogg, E., Schreier, E., \& Tananbaum, U.
1971, ApJ (Letters), 167, L67
\ms
\item{} Giacconi, R., 1974, in ``Astrophysics and Gravitation"
(Bruxelles: l'Universite de Bruxelles) p. 27
\ms
\item{} Gold, T. \& Hoyle, F. 1960, MNRAS, 120, 7
\ms
\item{} Gursky, H. \& Schreier, E. 1975, in Neutron Stars, Black Holes and
Binary
X-Ray Sources, ed. H. Gursky and R. Ruffini (D. Reidel Publishing
Company, Dordrecht), p. 175
\ms
\item{} Illarionov, A.F., \& Sunyaev, R.A. 1975, Astron. Astrophys., 39,
185.
\ms
\item{} Kaisig, M., Tajima, T., \& Lovelace, R.V.E. 1992, ApJ, 386, 83
\ms
\item{} Lamb, F.K., Pethick, C.J., \& Pines, D. 1973, ApJ, 184, 271
\ms
\item{} Lipunov, V.M. 1993, Astrophysics of Neutron Stars, (Berlin:
Springer-Verlag) chaps. 6,7
\ms
\item{} Lovelace, R.V.E., Berk, H.L., \& Contopoulos, J. 1991, ApJ, 379, 696
\ms
\item{} Lovelace, R.V.E., Mehanian, C., Mobarry, C.M., \& Sulkanen, M.E.
1986, ApJS, 62, 1
\ms
\item{} Lovelace, R.V.E., Romanova, M.M., \& Contopoulos, J. 1993, ApJ, 403,
158
\ms
\item{} Lovelace, R.V.E., Romanova, M.M., \& Newman, W.I. 1994, ApJ
(accepted)
\ms
\item{} Lynden-Bell, D., \& Boily, C. 1994, MNRAS, 267, 146
\ms
\item{} Lynden-Bell, D., \& Pringle, J.E. 1974, MNRAS, 168, 603
\ms
\item{} Mestel, L. 1968, MNRAS, 138, 359
\ms
\item{} Middleditch, J., \& Nelson, J. 1976, ApJ, 208, 567
\ms
\item{} Middleditch, J., Puetter, R.C., \& Pennypacker, C.R. 1985, ApJ,
292, 267
\ms
\item{} Nagase, F. 1981, Space Sci. Rev., 30, 395
\ms
\item{} Nagase, F., et al. 1984, ApJ, 280, 259
\ms
\item{} Nagase, F. 1989, Publ. Astron. Soc. Japan, 41, 1
\ms
\item{} Newman, W.I., Newman, A.L., \& Lovelace, R.V.E. 1992, ApJ, 392, 622
\ms
\item{} Parker, E.N. 1979, Cosmical Magnetic Fields (Oxford: Clarendon
Press) Chap. 17
\ms
\item{} Porter, L.A., Klimchuk, J.A., \& Sturrock, P.A. 1992, ApJ, 385, 738
\ms
\item{} Pringle, J.E., \& Rees, M.J. 1972, Astron. \& Astrophysics, 21,
1
\ms
\item{} Rappaport, S.A., \& Joss, P.C. 1977, Nature, 266, 683
\ms
\item{} Rappaport, S.A., \& Joss, P.C. 1983, in Accretion-Driven
Stellar X-Ray Sources, ed. W.H.G. Lewin and E.P.J. van den Heuvel
(Cambridge University Press, Cambridge), p.1
\ms
\item{} Schreier, E., Levinson, R., Gursky, H., Kellogg, E., Tananbaum,
H., \& Giacconi, R. 1972, ApJ (Letters), 172, L79
\ms
\item{} Schreier, E.J., \& Fabbiano, G. 1976, in X-Ray Binaries, NASA
SP-389 (National Technical Information Service, Springfield, Virginia),
p. 197
\ms
\item{} Shakura, N.I. 1973, Sov. Astron., 16, 756
\ms
\item{} Shakura, N.I., \& Sunyaev, R.A. 1973, Astron, \& Astrophys., 24, 337
\ms
\item{} Sheffer, E.K., Kopaeva, I.F.,Averintsev, M.B., Bisnovatyi-Kogan, G.S.,
Golynskaya, I.M., Gurin, L.S., D'yachkov, A.V., Zenchenko, V.M.,
Kurt, V.G., Mizyakina, T.A., Mironova, E.N., Sklyankin, V.A.,
Smirnov, A.S., Titarchuk, L.G., Shamolin, V.M., Shafer, E.Yu.,
Shmel'kin, A.A., \& Giovannelli, F. 1992, Sov. Astron.-AJ, 36, 41
\ms
\item{} Shvartsman, V.F. 1971, Soviet Astron.-AJ, 15, 377
\ms
\item{} Sturrock, P.A. 1991, ApJ, 380, 655
\ms
\item{} Tananbaum, H., Gursky, H., Kellogg, E.M., Levinson, R.,
Schreier, E., \& Giacconi, R. 1972, ApJ (Letters), 174, L143
\ms
\item{} Tjemkes, S.A., Ziuderwijk, E.J., \& Van Paradijs, J. 1986,
Astron. \& Astrophys., 154, 77
\ms
\item{} Wang, J.C.L., Lovelace, R.V.E., \& Sulkanen, M.E. 1990, ApJ,
355, 38

\vfill\eject
\baselineskip=20pt
\ctl{\bf{FIGURE CAPTIONS}}
\bs
\bs

\nid{{\it{Figure 1.~~}}} The figure shows the pulse period history for
pulsars Her X-1 (a) and GX 1+4 (b).  The first shows mainly spin-up
behavior with $\dot P = -3.35 \times 10^{-6}$s/yr.
Characteristic ``wavy" fluctuations of pulse period, with episodes of
spin-down, ${\dot P = 0}$, are alsoobserved.  The second pulsar shows
steady spin-up with ${\dot P = -2.55}$s/yr (GINGA observations, Bildsten 1993)
and then a change to spin down, ${\dot P = +1.5}$s/yr, and later $\dot P =
+2.51$s/yr (BATSE observations, Bildsten 1993).
\bs

\nid{{\it{Figure 2.}}} The figure shows the model of Ghosh and Lamb (1978).
 The existence of a broad transition zone with closed magnetic field
lines was suggested in this paper.
\bs

\nid{{\it{Figure 3.}}} A schematic drawing of the magnetic field
configuration considered in this work.  The magnetosphere consists of an
inner part, where the magnetic field lines are closed and an outer part
where the field lines are open.
\bs

\nid{{\it{Figure 4.}}} The figure shows two poloidal field lines used in
the derivation of equation (2).

\bs   \nid{{\it{Figure 5.}}} The figure shows perspective views of
illustrative field-line loops in the force-free corona of a Keplerian
disk.  In the bottom panel the difference in the angular displacement of
the foot points $\Dt \phi$ is zero, while the displacement increases
going to the middle panel $(\Dt \phi \approx 80^o)$, and then to the top
panel $(\Dt \phi \approx 320^o)$.  These field lines were determined
using the equations of Newman,
Newman and Lovelace (1992).

\bs
\nid{{\it{Figure 6.}}} The figure shows the radial dependences of
the main physical quantities for a case where there are a MHD outflows
from the $\pm z$ surfaces of the disk and where the pulsar spins up.  In
the plot of the disk's rotation rate, $r_{to}$ is the ``turn over
radius" where $d\om /dr = 0$.  In the plot of $F$, the radial scale is
different from the other plots.  In this plot $r_{wi}$ is the inner
radius of the MHD outflow.  We have taken $\ap = 0.1$, $\beta_\phi = 1$,
$D=1$, $\mu = 10^{30}$Gcm$^3$, $\dot M = 10^{17}$g/s, and $M = M_\odot$.
The values of $\om$ and $\om_K$ are in units of $(GM/r_o^3)^{1\ov 2}$,
$F_o = (GMr_o)^{1\ov 2}$, where $r_o = 0.91\times 10^8$cm.  The quantity
$T_{ss}$ is the Shakura and Sunyaev (1973) value.

\bs
\nid{{\it{Figure 7.}}} The figure shows the dependence of the turn-over
radius $r_{to}$ on $\mu$ for a case where $\ap = 0.1$, $\beta_\phi =
1,~D = 1$, $\dot M = 10^{17}$g/s, and $M = M_\odot$.

\bs
\nid{{\it{Figure 8.}}} The figure shows the radial dependences of
the main physical quantities for a case where there is {\it{no}}
outflow, but instead  magnetic braking of the accretion flow.  For the
case shown, the influx of angular momentum $\dot M F_\infty > 0$, and
thus the star spins up.  For this plot, $\ap = 0.1$, $D = 1$,
$\tau_{max} = 5$, $\mu = 10^{30}$G cm$^3$, $\dot M = 10^{17}$g/s, $M =
M_\odot$, $r_{cr} =0.91\times 10^8$cm, and $F_o = (GMr_{cr})^{1\ov 2}$.
For these
values $r_{cr}=r_{to}$.  The quantity $T_{ss}$
is the Shakura and Sunyaev (1973) value.

\bs
\nid{{\it{Figure 9.}}} The figure shows the
dependence of $F_\infty$ and $F_K = (GMr_{cr})^{1\ov 2}$ on the pulsar
period $P$
assuming $r_{cr} = r_{to}$ for ${\dot M}= 10^{17}$g/s and $4\times 10^{17}$
g/s.  The other parameters have been taken to be
$\ap = 0.1$, $ D = 1$, $\tau_{max} = 5$, and $M =M_\odot$.
Thus, each ${\dot M}=const.$ curve corresponds to different magnetic
moments as $\mu \propto P^{1.17}$.
$F_\infty$
and $F_K$ are measured in units of $(GMr_o)^{1\ov 2}$ with $r_o =
10^8$cm.
This plot allows a comparison of the angular momentum
influx to the star $(\dot M F_K)$ in the case of outflows and spin-up
$(r_{to} < r_{cr})$ with the case of magnetic braking where the
influx is $\dot M F_\infty (r_{to} > r_{rc})$.

\end